\documentclass[preprint,12pt]{elsarticle}
\usepackage{amsmath,amssymb,graphicx,booktabs}
\usepackage[dvipsnames]{xcolor}
\usepackage[hidelinks]{hyperref}
\usepackage{xurl}
\newcommand{\new}[1]{#1}

\journal{Journal of Information Security and Applications}

\begin{document}

\begin{frontmatter}

\title{ChronosAttack: Adversarial Tool Scheduling Attacks on LLM Agents}

\author[iut]{Masih Moafi}
\ead{m.moafi@el.iut.ac.ir}

\author[unb]{Arash Vashagh\corref{cor1}}
\ead{arash.vashagh@unb.ca}
\cortext[cor1]{Corresponding author.}

\affiliation[iut]{organization={Department of Electrical and Computer Engineering, Isfahan University of Technology},
            city={Isfahan},
            postcode={\new{84156-83111}},
            country={Iran}}

\affiliation[unb]{organization={Faculty of Computer Science, University of New Brunswick},
            city={Fredericton},
            postcode={E3B 5A3},
            state={NB},
            country={Canada}}

\begin{abstract}
We introduce ChronosAttack, an attack on large language model (LLM) agents that changes only when authentic tool results arrive. When an agent invokes several tools in parallel, the runtime may pass each result to the model as soon as it arrives. Tool completion times then determine the order in which the model reads the evidence. ChronosAttack exploits this behavior by delaying results without changing their contents or adding or removing observations. Checks that inspect output content alone therefore have no altered payload to flag. We formalize this delay-only threat model and consider a fixed attacker, which chooses a delivery order in advance, and an adaptive attacker, which tests orders using observed decisions. We distinguish \emph{order sensitivity}, the dependence of a decision on evidence order, from \emph{steering}, the attacker's ability to move a decision toward a chosen option. Using controlled replay on synthetic decision tasks, we evaluate both properties and compare defenses that change how agents receive or combine tool results. We assess these defenses by their resistance to steering and their effect on decisions in the absence of an attack.
\end{abstract}

\begin{keyword}
LLM agents \sep tool-use security \sep adversarial scheduling \sep order sensitivity \sep asynchronous agents \sep prompt injection
\end{keyword}

\end{frontmatter}


\section{Introduction}
Large language models (LLMs) are increasingly used as agents that call external tools, read what the tools return, and
act on the results~\cite{yao2023react, schick2023toolformer}. The model requests the calls. The receiving infrastructure executes the tools, collects their outputs, and assembles the next prompt. Early frameworks ran one call at a time~\cite{yao2023react}, whereas later systems run several calls at once to
reduce latency~\cite{kim2024llmcompiler} and either return the outputs in call order or pass each output to the model as
soon as its tool finishes~\cite{gim2024async}. When outputs arrive as tools finish, the infrastructure determines their order in the prompt.

Previous research on the security of LLM agents has primarily focused on the content they receive. Indirect prompt-injection attacks embed malicious instructions in content retrieved by agents~\cite{greshake2023indirect}. Existing benchmarks evaluate how frequently malicious tool outputs can manipulate an agent's behavior~\cite{debenedetti2024agentdojo}, while defense mechanisms aim to detect or isolate such content~\cite{debenedetti2026camel}. Separately, studies of order effects show that rearranging information within a prompt can change an LLM's response~\cite{lu2022fantastically,pezeshkpour2024order,liu2024lost}. These two lines of work leave a security question unanswered: when an asynchronous agent receives several authentic tool results, can an adversary steer its decision by controlling only when those results arrive? In this setting, the evidence is neither forged nor edited; its order is determined by the runtime's completion schedule.

ChronosAttack, introduced in our earlier preprint~\cite{vashagh2026chronos}, makes this scheduling behavior an explicit attack surface. An attacker delays tool results to change their delivery order while preserving every result's content. The agent can therefore make a different decision from the same evidence even though content integrity checks have nothing altered to detect. The attacker needs no access to the model, prompt, or tool-output contents; control over completion times, for instance through I/O or resource contention or the receiving infrastructure's completion queue, is sufficient.

For example, consider an agent choosing among three tree-planting programs based on four reports: equity, heat reduction, tree survival, and cost. The equity report favors one program, the heat report favors another, and the survival and cost reports favor the third. In our experiments, Qwen3 never selected the program favored by the equity report when the reports were presented in their natural order, with the equity report first. With the equity report last, the model selected that program in all 30 trials, despite receiving exactly the same report contents.

This journal version extends the original fixed-schedule attack in several ways. First, we formalize the delay-only threat model and distinguish \emph{order sensitivity} (whether a decision depends on result order) from \emph{steering} (whether an attacker can move it toward a chosen option). Second, we introduce an adaptive attacker that tests orders and selects one using observed decisions, exposing vulnerabilities a fixed order can miss. Third, we evaluate four defenses against both the attack and their own effects on clean decisions, including order-invariant synthesis, which summarizes each result independently before making the final decision. We also extend the empirical evaluation to additional models and decision tasks.

We evaluate ChronosAttack across six LLMs: GPT-5.6 Sol and GPT-5.6 Terra from OpenAI, Gemini 3.6 Flash from Google, Qwen3-235B from Alibaba, DeepSeek V4 Flash from DeepSeek, and Claude Sonnet 4.6 from Anthropic. Our evaluation includes six synthetic decision-making tasks: cloud backup, shipping partner selection, software release, microgrid management, tree planting, and water supply. Each task requires the agent to choose among several options without a single objectively correct answer.

Our results show that all four models tested on tree planting were sensitive to the order of tool results. However, the effectiveness of a particular attack order varied across models and depended on the decision context. The adaptive attacker also identified an effective ordering for a model that the fixed attacker failed to steer. Among the four defenses, only order-invariant synthesis consistently prevented substantial increases in the attacker's target selection.

Section~\ref{sec:background} reviews related work. Sections~\ref{sec:method} and~\ref{sec:setup} define the attack and evaluation. Sections~\ref{sec:results}--\ref{sec:conclusion} present the results, discuss their implications, and conclude.

\section{Related Work}\label{sec:background}
We relate ChronosAttack to content attacks on tool-using agents, order effects in language models, and timing attacks and defenses.

\subsection{Tool-Using Agents and Content Attacks}
Language models can learn when to call a tool and how to use its result~\cite{schick2023toolformer}, and agent
frameworks interleave such calls with reasoning~\cite{yao2023react}. Recent benchmarks place agents in environments that
change independently of the agent's actions~\cite{froger2026gaia2}, so the timing of observations becomes part of the
task. Our setting shares this asynchrony, with the difference that an attacker controls the timing.

In the content attacks considered here, the adversary changes what the agent reads. Injected content can lead an agent into harmful
actions~\cite{zhan2024injecagent, nakash2025footinthedoor}, and crafted tool descriptions and metadata can win tool
selection or carry hidden instructions~\cite{shi2026toolhijacker, wang2026mcptox}. Memory attacks insert records that
later steer the agent~\cite{dong2025memory, chen2024agentpoison}, and surveys organize these threats by where the
attacker acts and what the attacker can access~\cite{ferrag2026prompt, kim2026sok, vashagh2026survey}. In contrast, the
ChronosAttack adversary adds, alters, or removes nothing; the agent reads exactly the content that the tools produced.

\subsection{Order and Position Effects}
The sensitivity of language models to order is well documented. The order of in-context demonstrations can make the
difference between strong and chance performance, and an order that works well for one model does not transfer to
another~\cite{lu2022fantastically}. Reordering answer options~\cite{pezeshkpour2024order} or
premises~\cite{chen2024premise} changes accuracy, and relevant information is used less well in the middle of a long
context than at its ends~\cite{liu2024lost}. Similarly, LLM judges favor an answer because of its
position~\cite{li2024split}, and the size of this positional bias varies across models~\cite{schilcher2025positional}.

These studies fix one prompt, permute its parts, and measure a loss of accuracy or consistency. Our work differs in three
respects. First, the order is set by an attacker through tool completion times. Second, it unfolds over several agent
turns, so earlier evidence can shape provisional choices before later evidence arrives. Third, the quantity of interest
is how far the attacker can move the decision toward a chosen option.

\subsection{Timing Attacks and Defenses}
Timing has long been an attack surface in systems security. Time-of-check-to-time-of-use races, for example, exploit the
interval between a check and the action it guards~\cite{dean2004races}. Reordering authentic events without modifying
them has already been shown to defeat an adaptive monitor of electrocardiogram streams~\cite{vashagh2026conformalshift},
and removing context while leaving the target untouched can change the label that a classifier assigns to
it~\cite{vashagh2026cohorthijack}. In agents, timing appears in asynchronous monitoring of
actions~\cite{stickland2025async} and in work showing that the moment at which an injected instruction appears affects
its success~\cite{mou2026toolhazard}. In both cases, however, the attacker is the source of malicious behavior or
content, whereas in ChronosAttack only the timing is adversarial.

Ren et al. study how asynchronous delivery changes the search queries and additional evidence acquired by search
agents, and propose Confluent Evidence Acquisition to stabilize those paths~\cite{ren2026scheduler}. Their focus is
reproducible evidence acquisition and answer quality across schedules. We instead study an attacker who delays
unchanged tool results to steer a decision toward a chosen option, even when the agent performs no further retrieval.

Several agent defenses address malicious instructions or their effects. They separate instructions from data~\cite{chen2025struq}, re-execute
the agent to detect an injection's influence~\cite{zhu2025melon}, check that each action serves the user's
task~\cite{jia2025taskshield}, or attribute each tool call to its causes~\cite{he2026attriguard}. Like most defenses
against adversarial inputs~\cite{meymani2026defense}, each of them assumes a malicious input to find or isolate, and none
constrains the order in which authentic results reach the model. The defenses we evaluate act on that order instead; one
of them, voting over several orders, adapts the idea of aggregating answers over many
samples~\cite{wang2023selfconsistency}.

\subsection{Positioning of ChronosAttack}
ChronosAttack connects order effects in LLMs to an adversary's control over an asynchronous agent's runtime. Unlike
content attacks, it leaves every tool result authentic; unlike prompt-order studies, the adversary realizes an order
through delays in tool completion. The closest work on asynchronous search agents examines how timing changes which
evidence an agent subsequently acquires~\cite{ren2026scheduler}. Our setting holds the available evidence fixed and
asks whether its delivery order moves a final decision toward an attacker-selected target. This distinction motivates
measuring both general sensitivity to order and targeted steering, and evaluating defenses by the decisions they
produce as well as by whether they remove the attacker's control over order.

\section{Methodology}\label{sec:method}
We define the delay-only threat model, the fixed and adaptive attackers, and the metrics used to evaluate them. We then describe four defenses that change how the agent receives or combines tool results.

\subsection{Threat Model}
\begin{figure}[!t]\centering
\includegraphics[width=0.9\textwidth]{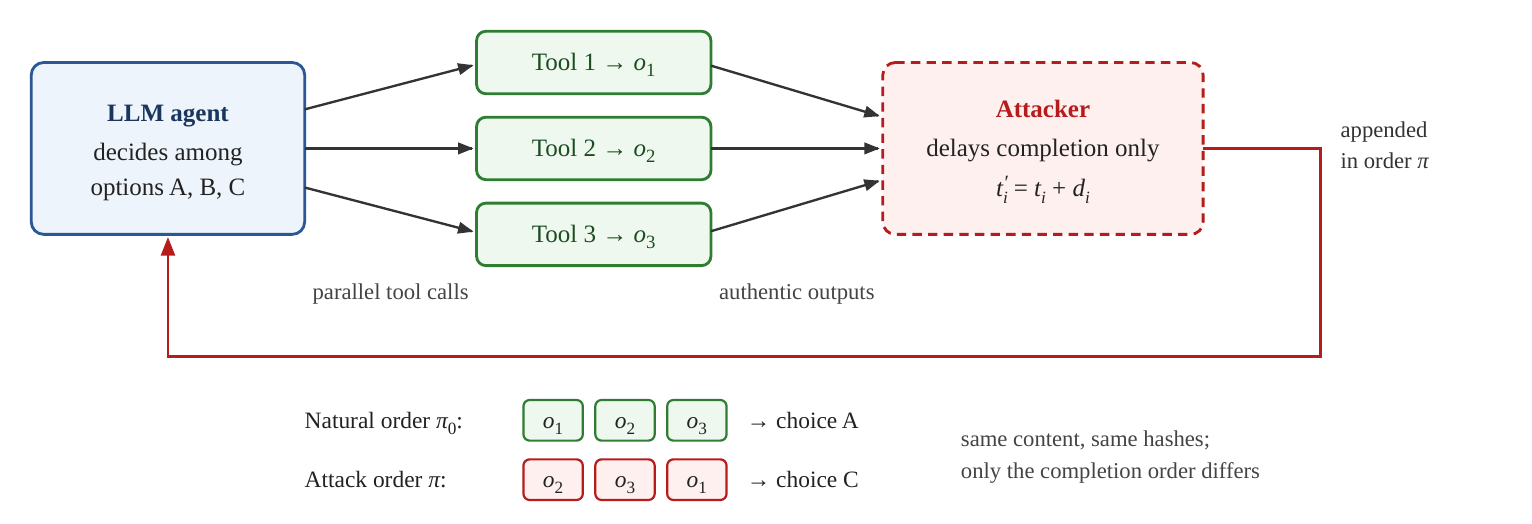}
\caption{The ChronosAttack threat model. The model requests parallel tool calls that run on the receiving infrastructure; the attacker delays tool
completions without altering their outputs. The same authentic outputs, appended to the prompt in a different order, can
lead to a different choice.}
\label{fig:threat}
\end{figure}

An agent must choose one option from a set $\mathcal{C}$. To do so, it requests $n$ tool calls in parallel. The receiving
infrastructure executes them, and tool $i$ finishes with an authentic observation $o_i$ at time $t_i$. Sorting the
completion times gives the natural order $\pi_0$, a permutation of $\{1,\dots,n\}$. The runtime appends the observations
to the model's context in a completion order $\pi$, so that $o_{\pi(k)}$ is the $k$-th observation the model reads;
without an attacker, $\pi=\pi_0$. Fig.~\ref{fig:threat} illustrates this setting.

The attacker is any component that can slow down tool completion or reorder the completion queue on the receiving
infrastructure. Examples include contention on local I/O or compute, a slow data source read by a tool, and the local
task scheduler itself. The attacker adds a non-negative delay $d_i$ to the completion of tool $i$, so that it finishes at
\begin{equation}\label{eq:delay}
t'_i = t_i + d_i, \qquad 0 \le d_i \le D,
\end{equation}
where $D$ is the largest delay the runtime tolerates before treating a call as failed. The delivered order $\pi$ is then
the order of the $t'_i$ in \eqref{eq:delay}. The attacker cannot alter, add, remove, duplicate, or accelerate any
observation. Hence, hashes and other integrity checks on the outputs pass. The attacker's goal is a target option
$c^\star\in\mathcal{C}$, chosen in advance.

We consider two types of attackers: fixed and adaptive. The fixed attacker selects a tool-result order $\pi_a$ in advance, based only on the available tools and decision options, and applies the same order across all models. In contrast, the adaptive attacker tests different orders, observes the agent's decisions, and selects the order $\pi^\star$ that most frequently leads to the target option $c^\star$. This attacker represents an adversary who can control tool-result scheduling and observe agent decisions over multiple interactions. Neither attacker can access the model weights or modify the prompt. We also assume agents do not receive tool completion timestamps, as these would provide timing information beyond the order of results.

\subsection{Attack Design}\label{sec:attack_design}

We consider two types of agents: stateful and stateless. A stateful agent processes one observation per turn, updates its internal state, and records a provisional decision after each observation. The agent retains its previous context when processing subsequent observations:
\begin{equation}\label{eq:stateful}
h_k = f(h_{k-1} \oplus o_{\pi(k)}),\ \ k=1,\dots,n; \quad y = g(h_n),
\end{equation}
where $h_0$ is the task prompt, $\oplus$ denotes concatenation, $f$ represents a model turn that updates the state and records a provisional decision, and $g$ represents the final turn that produces the decision $y\in\mathcal{C}$.

In contrast, a stateless agent receives all observations in a single model call, arranged according to their delivery order:
\begin{equation}\label{eq:stateless}
y = g\!\left(h_0 \oplus o_{\pi(1)} \oplus \dots \oplus o_{\pi(n)}\right).
\end{equation}
In both settings, tool results are provided to the model as user messages, consistent with how tool results are returned to Anthropic models.\footnote{\href{https://platform.claude.com/docs/en/agents-and-tools/tool-use/handle-tool-calls}{\nolinkurl{https://platform.claude.com/docs/en/agents-and-tools/tool-use/}\newline\nolinkurl{handle-tool-calls}}}
The agent is explicitly instructed not to interpret the arrival order of tool results as an indication of their importance, reliability, or relevance.

To manipulate the delivery order, the attacker introduces non-negative delays while preserving the original contents of all tool results. For a chosen ordering $\pi$, the attacker delays each tool completion until the preceding tool in that ordering has finished. The minimum delays required to achieve this ordering are calculated as
\begin{equation}\label{eq:schedule}
d_{\pi(k)} = \max\!\left(0,\; t'_{\pi(k-1)} + \epsilon - t_{\pi(k)}\right), \quad k=2,\dots,n,
\end{equation}
with $d_{\pi(1)}=0$, where $\epsilon$ is the smallest time gap required to distinguish consecutive completions. An ordering is feasible under the threat model only if every computed delay is at most $D$.

Because the agent does not receive completion timestamps, different feasible delay vectors that produce the same ordering result in identical model inputs. Consequently, ChronosAttack influences the agent's decisions through the order of tool results rather than their contents or the magnitude of the delays.

\subsection{Evaluation Metrics}\label{sec:metrics}

We evaluate ChronosAttack using metrics that capture both its ability to steer an agent toward a target option and the agent's general sensitivity to tool-result ordering.

To measure targeted steering, we evaluate each ordering $\pi$ over $R$ independent trials, producing decisions $y_\pi^{(1)},\dots,y_\pi^{(R)}$. The target-selection rate $\rho(\pi)$ and the effect of the fixed attack $\Delta$ are defined as
\begin{equation}\label{eq:rate}
\rho(\pi) = \frac{1}{R}\sum_{r=1}^{R} \mathbf{1}\!\left[y_\pi^{(r)} = c^\star\right], \qquad
\Delta = \rho(\pi_a) - \rho(\pi_0),
\end{equation}
where $\mathbf{1}[\cdot]$ is the indicator function, $\pi_a$ is the fixed attack order, and $\pi_0$ is the natural order. A positive observed value of $\Delta$ indicates an estimated increase in the probability of selecting the target option $c^\star$.

While the target-selection rate measures steering toward a particular option, it does not capture how sensitive the agent's decisions are to ordering in general. To evaluate order sensitivity, we consider an ordering $\Pi$ drawn uniformly from all $n!$ possible permutations and the corresponding decision $Y$. We quantify the dependence between ordering and decision using mutual information:
\begin{equation}\label{eq:mi}
I(\Pi;Y) = H(Y) - H(Y\mid\Pi),
\end{equation}
where $H$ denotes Shannon entropy in bits. A value of $I(\Pi;Y)=0$ indicates that the decision is statistically independent of the ordering, whereas a positive value indicates that the ordering provides information about the agent's decision.

Because LLM responses can vary across repeated trials, we also distinguish changes caused by ordering from those arising through sampling variability. Specifically, we compare the probability of obtaining different decisions under the same ordering with the probability of obtaining different decisions under distinct orderings:
\begin{equation}\label{eq:disagree}
\begin{aligned}
\delta_{\mathrm{w}} &= \Pr[Y\neq Y' \mid \Pi=\Pi'],\\
\delta_{\mathrm{b}} &= \Pr[Y\neq Y' \mid \Pi\neq\Pi'],
\end{aligned}
\end{equation}
where $Y'$ denotes the decision of an independent trial under ordering $\Pi'$. When $\delta_{\mathrm{b}}>\delta_{\mathrm{w}}$, decisions differ more frequently across distinct orderings than across repeated trials with the same ordering, providing additional evidence of order sensitivity.

Finally, we examine how the extent of reordering relates to attack effectiveness. We measure this using the inversion distance $\kappa(\pi,\pi_0)$, defined as the number of pairs of observations whose relative order differs between the delivered order $\pi$ and the natural order $\pi_0$. This metric allows us to investigate whether larger deviations from the natural ordering are associated with greater changes in the agent's decisions.

\begin{figure}[!t]\centering
\includegraphics[width=0.7\textwidth]{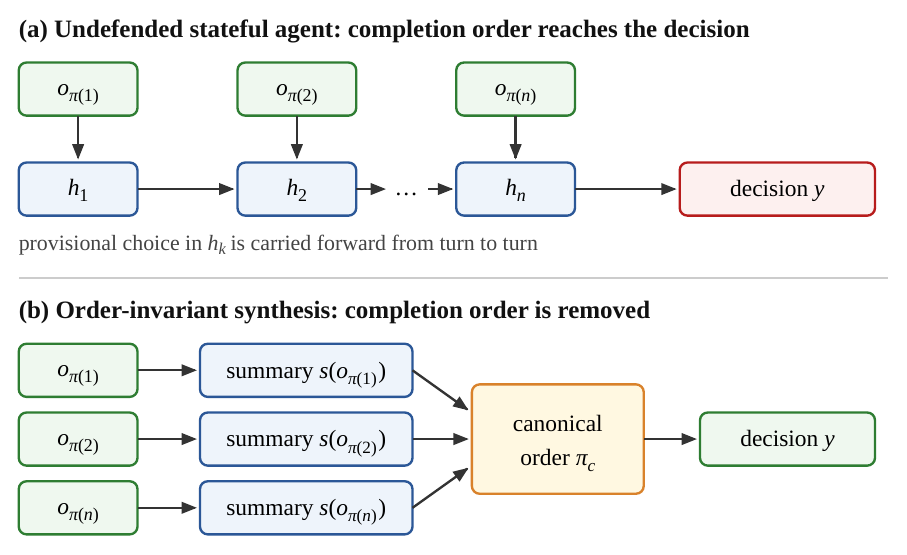}
\caption{Comparison of an undefended stateful agent and order-invariant synthesis. (a) The undefended agent processes tool results as they arrive, updating its state and provisional decision after each result. Consequently, changing the arrival order can affect its final decision. (b) In order-invariant synthesis, each tool result is summarized independently, and the summaries are presented in a fixed order before the final decision is made. This prevents the tool completion order from determining the order of information presented for the final decision.}
\label{fig:pipeline}
\end{figure}

\subsection{Order-Based Defenses}
The four defenses we evaluate all act on the order in which the model reads the results. The barrier waits for all $n$
results and presents them together in a fixed canonical order $\pi_c$. Randomized ordering presents the results in a
fresh random order for every decision, so that chance replaces the attacker's order. The order vote presents the results
in several orders, each as a single joint prompt, and returns the majority choice if it reaches a quorum and
abstains otherwise. The fourth defense, order-invariant synthesis, removes the order from the decision input altogether. Each observation is
first summarized on its own, in a separate call that sees no other observation, giving $s(o_i)$. The decision is then
made from the summaries in the canonical order:
\begin{equation}\label{eq:invariant}
y = g\!\left(h_0 \oplus s(o_{\pi_c(1)}) \oplus \dots \oplus s(o_{\pi_c(n)})\right).
\end{equation}
Since each summary is computed independently and the summaries are presented in a fixed order, the input of the final
call does not depend on $\pi$, as shown in Fig.~\ref{fig:pipeline}.

As a result, none of the four defenses lets the attacker choose the order that the model reads: the barrier and the vote
ignore $\pi$, randomized ordering replaces it, and \eqref{eq:invariant} removes it. The defenses differ, however, in what
the model reads instead, and that input can itself move the decision. We therefore judge each defense by whether the
target rate it produces stays at the natural rate $\rho(\pi_0)$ of \eqref{eq:rate}.

\section{Experimental Setup}\label{sec:setup}
We use six synthetic decision tasks, each with three options, and fix the target $c^\star$ before any model call. The
cloud-backup and shipping-partner tasks come from the preprint~\cite{vashagh2026chronos} and have three tools. The
software-release and microgrid tasks also have three tools and pose a comparable trade-off in new domains. The
tree-planting task has four tools, which report equity, heat reduction, tree survival, and cost, and each option is best
on at least one of them. The water-supply task repeats this four-tool structure with aquifer protection, drought yield,
lifecycle cost, and time to supply. In the four-tool tasks, the fixed order $\pi_a$ moves the report that favors the
target from first to last. For each prespecified experiment, the payloads, prompts, and candidate orders were frozen and hashed before collection; the adaptive order was selected from a completed 24-order panel and tested on fresh trials. Since no
option is correct in any task, the target rate \eqref{eq:rate} measures how far the decision can be moved, not how often
it is wrong.

The original study ran GPT-5.6 Sol, Gemini 3.6 Flash, DeepSeek V4 Flash, and Claude Sonnet 4.6, and we take its numbers
from the preprint. The new experiments ran on Kaggle Benchmarks with GPT-5.6 Terra, Gemini 3.6 Flash,
Qwen3-235B-A22B-Instruct-2507, and Claude Sonnet 4.6. Because the platform hosts neither GPT-5.6 Sol nor DeepSeek V4
Flash, the new evaluation uses Terra and Qwen rather than those models. We used the lowest accepted non-off reasoning
settings for Terra and Qwen (low) and Gemini (minimal). Claude ran with reasoning off. Its
minimal setting failed on every request, and with its low setting, one trial spent the whole answer cap on reasoning in
all three attempts, so that cell could not be completed. Every call used temperature 1 (as in the preprint), an answer cap of 256 tokens (enough for a choice and a short reason), a timeout of
30 seconds, and no seed. Every trial began a new chat. We replayed each schedule by presenting the same authentic
payloads to the model in the specified order. Arrival times and added delays were computed and logged, rather than
imposed on live tool executions; the models did not receive those timestamps. The experiments therefore isolate the
effect of delivered order, rather than measuring end-to-end exploitation of a deployed tool runtime.

For each model and task, we ran $R=30$ independent stateful trials under each of $\pi_0$ and $\pi_a$. Each condition used
new chats and model calls; matching repetition numbers were job identifiers, not a pairing mechanism. We also ran
ten stateless trials per order, ten trials with all observations presented at once and no completion order, 15 stateful
trials at each inversion distance, and 15 barrier trials. In the four-tool tasks, we additionally ran all 24 orders ten
times each, both statefully and as one joint prompt. These 24-order panels give $I(\Pi;Y)$ and the adaptive attacker's
order $\pi^\star$, which was then tested against each defense on 30 fresh trials per condition. The barrier used the
natural order as its canonical order, and the order vote used all 24 orders with a quorum of 16 votes. Natural
completion times in the four-tool tasks were 50, 100, 150, and 200~ms, and $\epsilon$ was 1~ms. Fixed-order effects were
tested with two-sided Fisher exact tests on the independent target-choice counts, with Holm correction~\cite{holm1979}
across the 16 new model-task comparisons in Table~\ref{tab:fixed}. We use 95\% Newcombe score intervals for differences
between independent proportions~\cite{newcombe1998independent}. The original preprint cells are descriptive here.
$I(\Pi;Y)$ was tested
against 999 shuffles of the orders within each panel, and differences between models in the effect of the order were
tested with Firth-penalized logistic regression~\cite{firth1993bias}. The stateful comparison of $\pi_0$ and $\pi_a$ is
the confirmatory endpoint; the adaptive attacker's order and every analysis chosen after seeing the data are
exploratory. For the new experiments, every call was logged before analysis, and every number reported below is recomputed from those logs.
The ablation analysis in Section~\ref{sec:ablation} examines sensitivity to these settings.

\section{Results}\label{sec:results}
We first evaluate the fixed attack's effectiveness. Table~\ref{tab:fixed} compares the target-selection rates under the natural and fixed attack orders, using 30 independent trials per order for each model-task combination. Part (a) presents the results from our original study~\cite{vashagh2026chronos} descriptively, while part (b) reports the new experiments. In part (b), bold marks differences significant under two-sided Fisher exact tests with Holm correction across the 16 cells ($p<0.05$). A
fixed order steered at least one model in every new task, but no fixed order steered every model. Qwen was steered in
all four new tasks, from 0\% to 100\% in three of them. Claude moved by 70 and 53 points in software release and
microgrid, Terra by 37 points in software release, and Gemini by 47 points in tree planting. Where a model already chose
the target, as Terra, Gemini, and Claude did in water supply, the attack had no room to act. The same order can also
push a model away from the target, as it pushed Gemini from 100\% to 7\% in cloud backup. Sequential state is not
required either. In the stateless setting \eqref{eq:stateless}, Claude's cloud-backup choice moved from 0 to 10 of 10
trials, and in tree planting Qwen was steered by one joint prompt. For Qwen, the effect is also not tied to the wording:
its steering was unchanged under two rewordings of the prompts and reports, and under two relabelings of the options it
moved from 0\% to 100\% in one and from 53\% to 77\% in the other. Part of Gemini's effect, in contrast, followed the
option letter, falling to 3\% under one relabeling.

\begin{table}[!htb]\centering\footnotesize
\caption{Fixed attack: stateful target rate (\%), natural $\to$ fixed order}\label{tab:fixed}
\setlength{\tabcolsep}{3pt}
\begin{tabular*}{\columnwidth}{@{\extracolsep{\fill}}lcc@{}}
\multicolumn{3}{@{}l}{(a) Original tasks and models}\\
\toprule
Model & Cloud backup & Shipping partner \\
\midrule
GPT-5.6 Sol & 0$\to$83 & 10$\to$67 \\
Gemini 3.6 & 100$\to$7 & 100$\to$80 \\
DeepSeek V4 & 7$\to$13 & 13$\to$20 \\
Claude 4.6 & 47$\to$93 & 0$\to$0 \\
\bottomrule
\end{tabular*}\par\medskip
\begin{tabular*}{\columnwidth}{@{\extracolsep{\fill}}lcccc@{}}
\multicolumn{5}{@{}l}{(b) New tasks and models}\\
\toprule
Model & Release & Microgrid & Tree & Water \\
\midrule
GPT-5.6 Terra & \textbf{57$\to$93} & 80$\to$90 & 3$\to$3 & 100$\to$100 \\
Gemini 3.6 & 93$\to$90 & 10$\to$30 & \textbf{47$\to$93} & 100$\to$100 \\
Qwen3 & \textbf{0$\to$100} & \textbf{0$\to$70} & \textbf{0$\to$100} & \textbf{0$\to$100} \\
Claude 4.6 & \textbf{20$\to$90} & \textbf{13$\to$67} & 93$\to$77 & 100$\to$100 \\
\bottomrule
\end{tabular*}
\end{table}

\begin{figure}[!t]
\centering
\includegraphics[width=\textwidth]{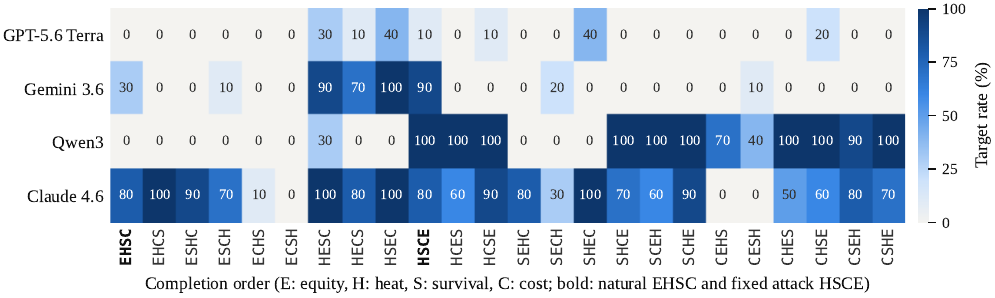}
\caption{Target-selection rates in the tree-planting task across all 24 possible tool-result orders, with ten stateful trials per order. Each ordering is represented by the sequence of four reports: equity (E), heat (H), survival (S), and cost (C). The natural order is EHSC, while the fixed attack order is HSCE.}
\label{fig:heatmap}
\end{figure}

Because all 24 orders were run in tree planting, this task also lets us separate order sensitivity
from steering. Table~\ref{tab:sens} shows that every model's choice depends on the order. $I(\Pi;Y)$ \eqref{eq:mi} ranges from
0.15 to 0.90 bits, each value with permutation $p=0.001$, and for every model, two decisions under different orders
disagree more often than two decisions under the same order \eqref{eq:disagree}. Fig.~\ref{fig:heatmap} gives the target
rate for each of the 24 orders, and the last column of Table~\ref{tab:sens} summarizes it: for every model, the target
rate ranges from 0\% under some orders to 40\% for Terra and 100\% for the other models under others. Steering toward
the target, however, needs room to move. Terra chooses the target in at most 40\% of trials under any order, and Claude
already chooses it under the natural order. The effect of the order also differs between models (penalized
likelihood-ratio $\chi^2(69)=327.9$, asymptotic $p=2.4\times10^{-35}$; exploratory, with uncertain calibration in this
sparse panel), and so does the observation that carries it. For Qwen, moving
the equity report from first to last raises the target rate by 100 points, whereas Gemini and Claude respond mostly to
the positions of the heat and cost reports.

\begin{table}[!htb]\centering\footnotesize
\caption{Tree planting: order sensitivity against sampling noise over all 24 orders}\label{tab:sens}
\setlength{\tabcolsep}{3pt}
\begin{tabular*}{\columnwidth}{@{\extracolsep{\fill}}lcccc@{}}
\toprule
Model & $I(\Pi;Y)$ (bits) & $\delta_{\mathrm{w}}$ & $\delta_{\mathrm{b}}$ & Range of $\rho$ (\%) \\
\midrule
GPT-5.6 Terra & 0.15 & 0.10 & 0.13 & 0--40 \\
Gemini 3.6 & 0.49 & 0.09 & 0.30 & 0--100 \\
Qwen3 & 0.90 & 0.07 & 0.52 & 0--100 \\
Claude 4.6 & 0.40 & 0.27 & 0.47 & 0--100 \\
\bottomrule
\end{tabular*}
\end{table}

\begin{figure}[!htb]\centering
\includegraphics[width=\columnwidth]{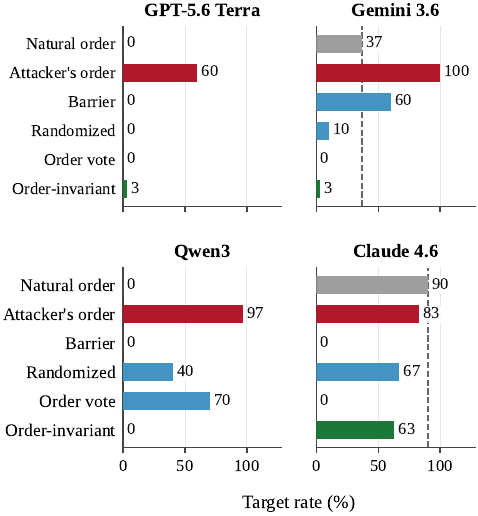}
\caption{Tree planting: target rate (\%) under each model's natural order, the adaptive attacker's order $\pi^\star$, and
$\pi^\star$ with each defense applied on its own. The dashed line marks the natural rate. $\pi^\star$ was chosen after
seeing each model's 24-order panel (exploratory).}
\label{fig:defenses}
\end{figure}

Since the fixed order is chosen without seeing any outcome, we next turn to the adaptive attacker. Fig.~\ref{fig:defenses}
shows its results together with those of the defenses, with 30 trials per condition. The adaptive attacker steered
Terra, which the fixed order had not moved, and moved Gemini and Qwen about as far as the fixed order had. On fresh
trials, its order $\pi^\star$ raised the target rate of Terra from 0\% to 60\%, of Gemini from 37\% to 100\%, and of Qwen
from 0\% to 97\%. For Claude, which already chose the target under the natural order, the search found no stronger
order. Because no defense lets the attacker choose the order, each defense gives the same input distribution against both attackers,
and the defenses differ only in where they leave the decision. Order-invariant synthesis \eqref{eq:invariant} removed
the fixed attack's effect on both steered models, from 27 to 0 of 30 trials for Gemini and from 30 to 0 for Qwen, and
against $\pi^\star$ it kept every model at or below its natural rate, apart from 3\% against 0\% for Terra. Each of the
other defenses, on the other hand, led at least one model to the target by itself. The barrier's single prompt in the
natural order did so for Gemini in 60\% of trials, against a natural rate of 37\%, and in 8 of 15 trials of the main
study. Randomized ordering did so for Qwen in 7 of 30 trials under the fixed attack and in 40\% here, since about half of
Qwen's 24 orders lead to the target, as Fig.~\ref{fig:heatmap} shows. The vote chose the target for Qwen in 9 of 10
order panels and in 70\% of decisions here, abstaining on the rest. Finally, order-invariant synthesis used five planned
model calls per decision for Terra and Claude and six for Gemini and Qwen, compared with five for the undefended agent
and 24 for the vote. Synthesis changed Gemini's clean choice distribution by 0.30 in total variation.

\subsection{Ablations and Hyperparameter Sensitivity}\label{sec:ablation}
We examine sensitivity to attack, model, and sampling settings.

\subsubsection{Attack Parameters}

Because models see neither timestamps nor delays, feasible delay vectors producing the same order give identical inputs. Equation~\eqref{eq:schedule} gives the minimum delays. Simulated completion times of 50, 100, 150, and 200~ms and of 2{,}350, 4{,}200, 6{,}050, and 7{,}900~ms produced byte-identical prompts under one order. For the fixed attack schedules, maximum computed delays were 136 and 66~ms in the original tasks and 151~ms in the four-tool tasks. These simulations do not establish a deployed runtime's tolerance $D$.

Table~\ref{tab:inversion} reports tree-planting target rates by inversion distance $\kappa(\pi,\pi_0)$, using 15 stateful trials of the lexicographically first order at each distance, chosen without outcomes. The fixed attack has distance 3. Qwen moves from 0\% at distances 0--3 to 100\% at 4--6; other models are nonmonotonic. Claude, for example, falls to 0\% at distance 3 and returns to 93\% at 6. The preprint likewise found that one inversion could be as effective as two~\cite{vashagh2026chronos}. Thus, distance alone does not characterize an attack order.

\begin{table}[!htb]\centering\footnotesize
\caption{Tree planting: target rate (\%) by inversion distance $\kappa(\pi,\pi_0)$}
\label{tab:inversion}
\setlength{\tabcolsep}{3pt}
\begin{tabular*}{\columnwidth}{@{\extracolsep{\fill}}lccccccc@{}}
\toprule
Model & 0 & 1 & 2 & 3 & 4 & 5 & 6 \\
\midrule
GPT-5.6 Terra & 7 & 0 & 0 & 7 & 0 & 0 & 0 \\
Gemini 3.6 & 40 & 0 & 40 & 0 & 0 & 0 & 0 \\
Qwen3 & 0 & 0 & 0 & 0 & 100 & 100 & 100 \\
Claude 4.6 & 87 & 93 & 67 & 0 & 67 & 80 & 93 \\
\bottomrule
\end{tabular*}
\end{table}

Removing each tree-planting report in turn leaves three observations and six orders. We tested every order ten times and averaged across the four removals, retaining the other payloads unchanged (Table~\ref{tab:three}). Excess order information is $I(\Pi;Y)$ minus its shuffled-order mean. With three reports, it falls near zero for Terra and Gemini; Gemini's highest target rate falls from 100\% to 32\%. Qwen remains more sensitive, while Claude shows a smaller decline. Removing evidence also changes the decision trade-off, and the four-report runs preceded the three-report runs by a month, so this comparison does not isolate observation count.

\begin{table}[!htb]\centering\footnotesize
\caption{Tree planting: four against three observations}\label{tab:three}
\setlength{\tabcolsep}{3pt}
\begin{tabular*}{\columnwidth}{@{\extracolsep{\fill}}lcccc@{}}
\toprule
 & \multicolumn{2}{c}{Excess $I(\Pi;Y)$ (bits)} & \multicolumn{2}{c}{Highest $\rho$ (\%)} \\
Model & Four & Three & Four & Three \\
\midrule
GPT-5.6 Terra & 0.07 & 0.00 & 40 & 25 \\
Gemini 3.6 & 0.40 & 0.02 & 100 & 32 \\
Qwen3 & 0.80 & 0.48 & 100 & 75 \\
Claude 4.6 & 0.32 & 0.20 & 100 & 82 \\
\bottomrule
\end{tabular*}
\end{table}

\subsubsection{Model and Sampling Parameters}

Table~\ref{tab:reasoning} compares Terra with reasoning off and low (the main setting), using 30 trials per order and the 24-order panels. The attack increased target rates by 83 and 53 points in release and microgrid with reasoning off, versus 37 and 10 with low reasoning, partly because the low-reasoning natural rates were higher. Tree-planting target rates remained near zero, and model-by-order statistics were similar ($\chi^2(69)=319.7$ versus 327.9). The runs were a month apart, so the comparison is descriptive.

\begin{table}[!htb]\centering\footnotesize
\caption{GPT-5.6 Terra: reasoning off against low}
\label{tab:reasoning}
\setlength{\tabcolsep}{3pt}
\begin{tabular*}{\columnwidth}{@{\extracolsep{\fill}}lcccc@{}}
\toprule
Setting & Release (\%) & Microgrid (\%) & Tree (\%) & Tree $I(\Pi;Y)$ \\
\midrule
Off & 17$\to$100 & 33$\to$87 & 0$\to$7 & 0.09 \\
Low & 57$\to$93 & 80$\to$90 & 3$\to$3 & 0.15 \\
\bottomrule
\end{tabular*}
\end{table}

With $R=30$ independent trials per order, every Holm-significant effect among the 16 new comparisons was at least 37 points. Gemini's 20-point microgrid increase was not significant (95\% independent-proportions interval: $-1$ to $+39$ points). An order selected from five of ten trials per permutation raised the target rate on the remaining five by 100 points for Qwen and 60 for Gemini. This split-half check is descriptive; Fig.~\ref{fig:defenses} reports 30 fresh trials per condition. The 256-token answer cap also constrained Claude's reasoning settings (Section~\ref{sec:setup}).

\section{Discussion and Future Directions}\label{sec:discussion_future_direction}
These results support reporting order sensitivity and steering as separate properties. Every model in
tree planting changed its decision with the order of the same evidence, yet a given order steered only some models in
some tasks, and the direction of the shift differed between models. A single attack success rate, measured with one
order on one model, would describe neither property. An evaluation is therefore more informative when it reports, for
each model and task, the order information together with the effect of a fixed order and of the order found by search.

Steering also depends on the model's natural choice. In the water-supply task, three of the four models already selected the target under the natural order, leaving no room to increase its selection rate. The ablations in Section~\ref{sec:ablation} also show that removing one report substantially reduced order sensitivity for Terra and Gemini in tree planting.

Most models changed their provisional decisions approximately once per trial, while Qwen did so approximately twice. These changes occurred most frequently after the second observation, and the final decision rarely differed from the last provisional choice. These trajectories suggest that evidence received at some stages may influence the final choice more than evidence received at others. Gemini's sensitivity to option labels also suggests that ordering effects may depend on how the choices are presented.

The study uses synthetic decisions without an objectively correct option and replays tool-result orders rather than delaying live tools. It establishes decision sensitivity under controlled evidence but does not measure task accuracy or whether an attacker can induce the required delays in a deployed runtime.

Experiments with agents that use external services or execute complex workflows could test whether the observed effects persist under realistic network delays and resource constraints.

Adaptive attacks on previously unseen agents also need study. Measuring how many observations of agent decisions an attacker needs to find an effective order would help assess attacks with limited feedback.

An adaptive defense could summarize each tool result independently and request further analysis only when the evidence conflicts or the decision remains uncertain. Whether this reduces cost while preserving resistance to scheduling attacks should be tested together with latency and clean decision quality.

\section{Conclusion}\label{sec:conclusion}
ChronosAttack shows that an agent's final choice can depend on the delivery order of authentic tool results. The delay-only threat model separates this scheduling attack from content manipulation and distinguishes general order sensitivity from steering toward a chosen option. In the evaluated tasks, a model that resisted one fixed order could still be steered by an order found through search. Removing the attacker's control over delivery order also did not guarantee preservation of the agent's clean decisions. Order-invariant synthesis most consistently prevented targeted steering in the tree-planting experiments, but changed some models' choices without an attack. These controlled replay results motivate evaluation in deployed agents, where the attacker's ability to induce the necessary delays remains to be established.

\new{\section*{Declaration of generative AI and AI-assisted technologies in the manuscript preparation process}
During the preparation of this work, the authors used OpenAI Codex and Anthropic Claude to assist with implementing the experiment and analysis code, language editing, document formatting, and checks of wording, references, and numerical consistency. After using these tools, the authors reviewed and edited the content as needed and take full responsibility for the content of the published article. The research use of language models is described in Section~\ref{sec:setup}.}

\new{\section*{CRediT authorship contribution statement}
\textbf{Masih Moafi:} Methodology, Software, Validation, Formal analysis, Investigation, Data curation, Visualization, Writing -- review \& editing. \textbf{Arash Vashagh:} Conceptualization, Methodology, Software, Investigation, Writing -- original draft, Writing -- review \& editing, Supervision, Project administration.}

\new{\section*{Declaration of competing interest}
The authors declare that they have no known competing financial interests or personal relationships that could have appeared to influence the work reported in this paper.}

\new{\section*{Funding}
This research did not receive any specific grant from funding agencies in the public, commercial, or not-for-profit sectors.}

\section*{Data availability}
The source code, task definitions, prompts, and raw model outputs supporting the findings of this study are publicly available at \url{https://github.com/arashVsh/ChronosAttack} in the \texttt{journal\_extension} folder. The repository includes a README file with instructions for reproducing the reported results from the raw outputs without additional model calls and for rerunning the experiments. Users who rerun the experiments must provide their own model access.

\bibliographystyle{elsarticle-num}
\bibliography{refs}

\end{document}